\documentclass[preprint,showpacs,preprintnumbers,amsmath,amssymb,nofootinbib]{revtex4-1}
\usepackage[dvips]{graphicx}
\usepackage{graphicx}
\usepackage{amsfonts}
\usepackage{bm}
\usepackage{amsmath}
\usepackage{amssymb}
\usepackage{color}
\usepackage[all]{xy}

\def\be{\begin{equation}}
\def\ee{\end{equation}}
\def\bea{\begin{eqnarray}}
\def\eea{\end{eqnarray}}

\begin{document}

\title{Thermodynamics of the  FLRW apparent horizon in Einstein-Gauss-Bonnet gravity}

\author{Luis M. S\'anchez$^1$ and Hernando Quevedo$^{2,3}$   }
\email{lsanchezh@uaemex.mx,quevedo@nucleares.unam.mx}
\affiliation{$^1$Facultad de Ciencias, 
	Universidad Aut\'onoma del Estado de M\'exico, Campus Universitario “El Cerrillo, Piedras Blancas”, Carretera Toluca-Ixtlahuaca km 15.5, C.P. 50200 Estado de México, Mexico}
\affiliation{$^2$Instituto de Ciencias Nucleares, 
	Universidad Nacional Aut\'onoma de M\'exico, 
	AP 70543, M\'exico, DF 04510, Mexico}
\affiliation{$^3$ 
	Dipartimento di Fisica and ICRA, Universit\`a di Roma ``La Sapienza", Piazzale Aldo Moro 5, I-00185 Roma, Italy}

\begin{abstract}
We analyze the thermodynamic properties of the apparent horizon of Friedmann-Lema\^itre-Roberson-Walker (FLRW) spacetimes in Einstein-Gauss-Bonnet gravity. We use the generalized definition of entropy for gravity theories in higher dimensions to determine the main thermodynamic variables and to compare their behavior with the corresponding quantities in Einstein's theory, emphasizing the role of the Gauss-Bonnet coupling constant and the dimension number. By imposing the validity of the laws of thermodynamics, we show that the apparent horizon can be interpreted thermodynamically as a dark energy fluid, independently of the coupling constant and the dimension number. Using the response functions, we determine the adiabatic index and the number of thermally accesible degrees of freedom of the apparent horizon and argue that this leads to a discretization of the Gauss-Bonnet coupling constant.

\end{abstract}


\maketitle

\section{Introduction}
\label{sec:int}

The idea that gravity could be interpreted as an emergent theory was first proposed by A.D. Sacharov in 1967 in \cite{sakharov1967vacuum}. It was suggested that spacetime is the result of a mean-field approximation of an unknown theory of underlying degrees of freedom, resembling the way how thermodynamics emerges from statistical molecular physics. This suggestion opened the possibility of understanding gravity from a thermodynamic point of view. For a recent review of different approaches, see \cite{padmanabhan2010thermodynamical}. 

The first evidence for the connection between gravity and thermodynamics appeared with black hole thermodynamics as proposed by J. Bekenstein in \cite{Bekenstein:1972tm}. It turns out that the entropy of a black hole is related to the surface of the event horizon, and the temperature is defined in terms of the surface gravity. Thus, a black hole can be understood as a classical thermodynamic system whose fundamental equation is determined by the Bekenstein-Hawking formula that relates the entropy with the independent physical parameters of the black hole, such as the mass and angular momentum. As a result, the mass satisfies the first law of thermodynamics and the entropy behaves as in a classical thermodynamic system. This approach represents the
fundamentals of black hole thermodynamics (for a recent review, see, for instance, \cite{carlip2014black}). 
 
However,  black holes represent only a particular case of spacetimes that can be interpreted as thermodynamic systems. In fact, to any solution of general relativity that possesses a horizon, one can associate a thermodynamic entropy that is proportional to the area of the corresponding horizon. In particular, the Rindler space and the de-Sitter universe are characterized by the existence of horizons that carry the entropy of the entire spacetime. The question arises whether an arbitrary spacetime that is locally equivalent to a Rindler space can be endowed with an entropy which is proportional to the local Rindler horizon and satisfies the laws of thermodynamics. If so, it would imply that there exists a geometric representation of the thermodynamic laws, in particular, the Clausius relationship would imply the dynamical equations of the spacetime, i.e.,
\be
T\nabla S = Q \Longrightarrow R_{\mu\nu} - \frac{1}{2} R g_{\mu\nu} + \Lambda g_{\mu\nu} = 4\pi T_{\mu\nu},
\ee
as proposed by T. Jacobson in \cite{jacobson1995thermodynamics}.

This idea has been applied in the case of cosmological models, where it was proved that the apparent horizon of a Friedmann-Lema\^itre-Robertson-Walker (FLRW) can be equipped with an entropy proportional to the area of the horizon, from which the Friedmann equations are derived \cite{cai2005first}.
Following this approach, in a previous work \cite{sanchez2023thermodynamics}, we assumed that the FLRW apparent horizon itself is a thermodynamic system that can be handled by using the standard relations of classical thermodynamics. As a result, we proved that, in fact, the apparent horizon can be characterized by a fundamental equation, from which all the properties of the system are derived. In particular, we found that the apparent horizon satisfies a dark-energy-like equation of state and satisfies the corresponding laws of thermodynamics when considered as a system surrounded by the fluids of the standard cosmological model. Other aspects of the application of thermodynamics in cosmology have been analyzed in \cite{nojiri2025modified,odintsov2025natural,paul2025origin}.

To find out if the above result can be generalized to other gravity theories, in this work, we study the FLRW spacetime in the Einstein-Gauss-Bonnet (EGB) gravity theory. Several interesting thermodynamic aspects of the FLRW spacetime in the EGB theory have been analyzed in \cite{saavedra2025thermodynamic,paul2024gauss}. 
In this work, we will show that, in fact, the FLRW apparent horizon can be handled as a thermodynamic system. The thermodynamic variables and the response functions turn out to depend explicitly on the Gauss-Bonnet (GB) coupling constant and on the number of dimensions. Remarkably, the equation of state of the horizon turns out to be independent of the details of the EGB theory.

This work is organized as follows. In Sec. \ref{sec:review}, we review the main concepts of the EGB theory and present and analyze the field equations, which describe the evolution of the FLRW cosmological model. In Sec. \ref{sec:flrw}, we consider the properties of the apparent horizon in the FLRW spacetime, and show that in the EGB theory, the horizon radius is always larger than in Einstein's gravity. In Sec. \ref{sec:laws}, we derive the explicit form of the main thermodynamic variables and  explore the conditions that the apparent horizon must satisfy for the laws of thermodynamics to be fulfilled. The consequences of this assumption are presented in Sec. \ref{sec:eos}, where it is shown that the apparent horizon satisfies a dark-energy-like equation of state, implying that the thermodynamic evolution of the horizon is isentropic. We argue that this result can lead to a discretization of the GB coupling constant. Finally, in Sec. \ref{sec:con}, we summarize and discuss our results.

%%%%%%%%%%%%%%%%%%%%%%%%%%%%%%%%%%%%%%%%%%
\section{The FLRW spacetime in EGB gravity}
\label{sec:review}

The action of EGB gravity theory can be written as \cite{cai2005first}
\begin{equation}
     S= \frac{1}{16 \pi} \int d^{n+1}x\sqrt{-g}\left( R - 2\Lambda + \alpha_0 \mathcal{G} \right) + S_m,
 \end{equation}
 where $\alpha_0$ is the GB coupling constant,  $\Lambda$ is the cosmological constant, $S_m$ represents the action of the matter, and  $\mathcal{G}$  is the so-called GB term given by
\begin{equation}
\mathcal{G} = R^2 - 4R_{\mu\nu}R^{\mu\nu} + R_{\mu\nu\alpha\beta}R^{\mu\nu\alpha\beta}.
\end{equation}
In four dimensions, the GB term is purely topological and, therefore, it does not contribute to the dynamics. What distinguishes the EGB gravity is that, even though the action involves higher-order curvature terms, the equations of motion derived from it  contain only second-order derivatives of the metric. Indeed, it can be shown that by varying the action, one obtains the equations of motion
\begin{equation}
R_{\mu\nu} - \frac{1}{2}R g_{\mu\nu} + \Lambda g_{\mu\nu} - \alpha_0 \left( \frac{1}{2}\mathcal{G} g_{\mu\nu} - 2RR_{\mu\nu} + 4 R_{\mu\alpha} R^{\alpha}_{\ \nu} + 4 R_{\alpha \beta} R^{\alpha\ \beta }_{\ \mu \ \nu} - 2R_{\mu\alpha\beta\gamma}R_{\nu}^{\ \alpha\beta\gamma} \right) = 8 \pi T_{\mu\nu}.
\end{equation}

In this wirk, we will consider the FLRW cosmological model, which  is described by the metric 
\be
ds^2 = h_{ab}dx^a dx^b + R^2 d\Omega_{n-1}^{2}\ ,\ \ h_{ab}= {\rm diag}\left(-1,\frac{a^2(t)}{1-kr^2}\right) ,
\ee
where the index $a=0,1$, $d\Omega_{n-1}^{2}$  is the line element of a unit $(n-1)$  dimensional sphere, $k=0,\pm 1$ is the curvature constant, 
the coordinates $x^a$ are defined as $x^0 = t$ and $x^1 =r$, and we introduce the notation  $R=a(t) r$. Moreover, the gravitational source of the FLRW spacetime is assumed to be determined by the energy-momentum tensor of a perfect fluid
\begin{equation}
T_{\mu\nu}= (\rho +p ) u_{\mu}u_{\nu} + pg_{\mu\nu}. 
\end{equation}

Then, the EGB field equations for the FLRW metric reduce to the Friedmann equations
\cite{cai2005first}
\begin{equation}
\left[1 + 2{\alpha}\left( H^{2}+\frac{k}{a^{2}}\right) \right] \left( \dot{H}-\frac{k}{a^{2}}\right) = -\frac{8 \pi G}{n-1}\left( \rho + p\right),
\label{firstfriedmaneq}
\end{equation}
\begin{equation}
 H^{2}+\frac{k}{a^{2}} +{\alpha} \left( H^{2}+\frac{k}{a^{2}}\right)^{2} = \frac{16 \pi }{n(n-1)} \rho,
 \label{secondFridmaneq}
\end{equation}
where $H=\frac{\dot a}{a}$ is the Hubble parameter and ${\alpha} =(n-2)(n-3)\alpha_0$. 
Notice that when $\alpha_0 =0$, the Friedmann equations of general relativity are recovered. In this work, we will limit ourselves to the case with no cosmological constant to explore the explicit contribution of the Gauss-Bonnet term. 

For the investigation of the evolution equations, it is convenient to consider the conservation law of the energy-momentum tensor
\be
\dot \rho + nH(\rho+p)=0\ , 
\label{claw}
\ee
which also follows from the Friedmann equations. In the case of a barotropic perfect fluid with equation of state $p=w\rho$, the conservation law (\ref{claw}) can be integrated explicitly and yields
\be
\rho = \rho_0 \left( \frac{1+z}{a_0}\right)^{n(1+w)},
\ee
where $\rho_0$ is an integration constant and, for later use, we have introduced the redshit parameter $z$ using the definition relation 
\be
z = \frac{a_0}{a} - 1,
\ee
with $a_0=const.$

%%%%%%%%%%%%%%%%%%%%%%%%%%%%%%%%%%%%%%%%%%%%%%%
%%%%%%%%%%%%%%%%%%%%%%%%%%%%%%%%%%%%%%%%%%%%%%%%

\section{The FLRW apparent horizon }
\label{sec:flrw}

An important feature of the FLRW spacetime is the existence of a marginally trapped null surface, known as the apparent horizon, characterized by a vanishing expansion. This surface satisfies the condition \cite{faraoni2011cosmological,faraoni2019cosmological}
\be 
h^{ab}\frac{\partial R}{\partial x^a}\frac{\partial R}{\partial x^b} =0\ ,
\ee 
%It is worth noting that this condition depends solely on the metric, and therefore remains the same in both General Relativity and Gauss-Bonnet gravity. 
whose solution reads  
\begin{equation}
    R_h = \frac{1}{\sqrt{
		H^2 + \frac{k}{a^2} } }
\label{hor}
\end{equation}
and determines the radius of the apparent horizon. Using the Friedmann equations for a barotropic perfect fluid, $p=w\rho$, the derivative of the apparent horizon radius can be expressed as 
\begin{equation}
\dot{R}_h = -R^{3}_h H \left( \dot{H}-\frac{k}{a^{2}} \right) = {\frac {8 \pi \,\rho  \left( \omega+1 \right) H R_h^{3}}{n-1} \left( 1+2\,{\frac {
\alpha}{ R_h^{2}}} \right) ^{-1}},
\label{dotR}
\end{equation}
which reflects the dynamic nature of the apparent horizon.

In the context of a dynamical spacetime, the apparent horizon has been proposed to act as a causal horizon, with an associated gravitational entropy and surface gravity  \cite{hayward1994general, hayward1996gravitational,bak2000cosmic,hayward1998unified,faraoni2011cosmological,melia2018apparent}. As a result, it is possible to assume that the apparent horizon behaves as a thermodynamic system, to which one can associate the temperature 
\be
T_h= \frac{\vert\kappa\vert}{2\pi},
\ee
where the surfave gravity $\kappa$
 is defined as\cite{cai2005first,hayward1998unified, binetruy2015apparent}
\begin{equation}
\kappa = \frac{1}{2 \sqrt{-h}} \frac{\partial}{\partial x^{a}} \left(\sqrt{-h} h^{ab} \frac{\partial R_h}{\partial x^{b}} \right) = -\frac{1}{R_h } \left(1- \frac{1}{2} 
\frac{ \dot{R}_h}{ H R_h } \right)\ .
\end{equation}
Here, $h$ denotes the determinant of the metric $h_{ab}$. Then, the temperature of the apparent horizon becomes 
\begin{equation}
T_h = \frac{1}{2 \pi R_h } \left|1- \frac{1}{2}\frac{ \dot{R}_h}{ H R_h} \right|\ .
\label{temp}
\end{equation}

Notice that in the limiting case of a non-dynamical radius or a very slowly changing apparent horizon with $\frac{\dot{R}_h}{2 H R_h} \ll 1$, the temperature reduces to $
T_h=\frac{1}{2\pi R_h}$, an expression that resembles the temperature of a spherically symmetric black hole with horizon radius $R_h$.

Furthermore, considering the apparent horizon as a classical thermodynamic system, one can introduce the first law of thermodynamics in the form 
\cite{hayward1998unified,hayward1994general,hayward1996gravitational}
\be
dE = A \Psi + W dV 
\label{uflaw1}
\ee
which is known as Hayward’s unified first law. 
Here, $A=n\Omega_n R^{n-1}$ corresponds to the surface area and $V=\Omega_n R^{n}$ to the volume of an $n-$dimensional space with radius $R$. The factor $\Omega_n =\frac{\pi^{n/2}}{\Gamma(n/2+1)}$ represents the volume of a unit sphere in $n$ dimensions.

In this context, $\Psi$ is the energy flux 
\be
\Psi = \Psi_ a dx^a = \left(T_a^{\ b}\frac{\partial R}{\partial x^b} + W \frac{\partial R}{\partial x^a}\right)dx^a \ ,
\ee
and $W$ is the work density
\be
W = - \frac{1}{2} T^{ab}h_{ab}\ .
\ee

In the case of the FLRW spacetime described above, we obtain at the apparent horizon 
\begin{equation}
\Psi_ a = \left( -\frac{1}{2}(\rho +p)H R,\frac{1}{2}(\rho +p)a\right)
\end{equation}
and 
\be
W=\frac{1}{2} (\rho-p)\ .
\label{work0}
\ee
Note that the energy flux vector and the work density depend solely on the energy-momentum tensor and the metric in the $(t,r)$ plane. However, both quantities are independent of the field equations. 
Therefore, in EGB gravity, the thermodynamic quantities $\Psi_a$ and $W$ coincide with those of general relativity. The same holds for the surface gravity, as it is defined within the $(t,r)$ plane and depends exclusively on the metric $h_{ab}$.

In contrast, the entropy does not exhibit a similar behavior, as it is a quantity that explicitly depends on the gravitational theory under consideration. In theories beyond general relativity, the entropy is no longer proportional to one-quarter of the horizon area. Specifically, in the EGB theory, the entropy of the horizon is given by \cite{cai2005first}
\begin{equation}
S= \frac{A}{4}\left[1+ \frac{2{\alpha}(n-1)}{(n-3)R_h^{2}} \right] = \frac{n\Omega_n {R_h}^{n-1}}{4} \left[1+ \frac{2{\alpha}(n-1)}{(n-3)R_h^{2}} \right] ,
\label{GBEntropy}
\end{equation}
where $A=n\Omega_n R^{n-1}_h$ is the area of the apparent horizon. 
Since in this framework, entropy should still be  proportional to the area, which is a positive quantity, we may demand the positivity of the above expression. Then, it is easy to show that this condition implies that 
\be
R_h^2+ 2\alpha > \frac{2}{n-1} R_h^2 > 0. 
\label{condR}
\ee
In the following sections, we will use this inequality to estimate the behavior of certain thermodynamic variables.

%%%%%%%%%%%%%%%%%%%%%%%%%%%%%%%%%%%%%%

\subsection{Analysis of the apparent horizon radius}

The second Friedmann equation (\ref{secondFridmaneq}) can be expressed in terms of the apparent horizon radius (\ref{hor}) as
\begin{equation}
   {\frac {1}{{R}_h^{2}}}+{\frac {{\alpha}}{{R}_h^{4}}}=\beta \rho,
\end{equation}
where 
\be
{\alpha} =(n-1)(n-2)\alpha_0 ,
\quad \beta = \frac{16 \pi }{n(n-1)}
.
\ee
Solving the algebraic quadratic equation for the radius of the horizon, we find that the only physically viable solution is
\begin{equation}
     R_h^2 =  {{\frac {1+\sqrt {1+ 4 \alpha \beta \rho }}{2 \beta\rho }}},
     \label{GBrad}
\end{equation}
which in the limiting case $\alpha =0$ and  $n=3$ reduces to the corresponding value in Einstein gravity \cite{sanchez2023thermodynamics}
\begin{equation}
    R_h^E =\sqrt{\frac{3}{8 \pi \rho}}\ .
\end{equation}

From the expression for the horizon radius (\ref{GBrad}), we see that the GB coupling constant $\alpha_0$ cannot be chosen arbitrarily, but it must satisfy the condition 
\be
\alpha_0 \geq - \frac{n}{64\pi (n-2) \rho }, 
\label{cond1}
\ee
for the radius to be a real quantity. This condition implies that the EGB theory for negative values of the coupling constant $\alpha_0$ is restricted by the energy density of the universe\footnote{Remarkably, this resembles the behavior of some quantum field theories such as quantum chromodynamics \cite{Schwartz_2013}.}. Consequently, at higher energies the coupling constant decreases, and the apparent horizon in the EGB theory becomes effectively equivalent to that of Einstein's theory.

From Eq.(\ref{GBrad}), it follows that
\be
R_h \geq \sqrt{\frac{n(n-1)}{32\pi \rho}},
\label{GBradmin}
\ee
where, according to Eq.(\ref{cond1}), the minimum value is reached for $\alpha_0 = -\frac{n}{64\pi (n-2)\rho }$. Furthermore, from the above equation for $n=4$, we obtain 
\be
R_h(n=4) \geq  \sqrt{\frac{3}{8\pi \rho}} = R_h^E,
\ee
which implies that for $n\geq 4$ the radius of the FLRW apparent horizon in the EGB theory is always larger than in Einstein's theory.

\begin{figure}
\includegraphics[scale=0.5]{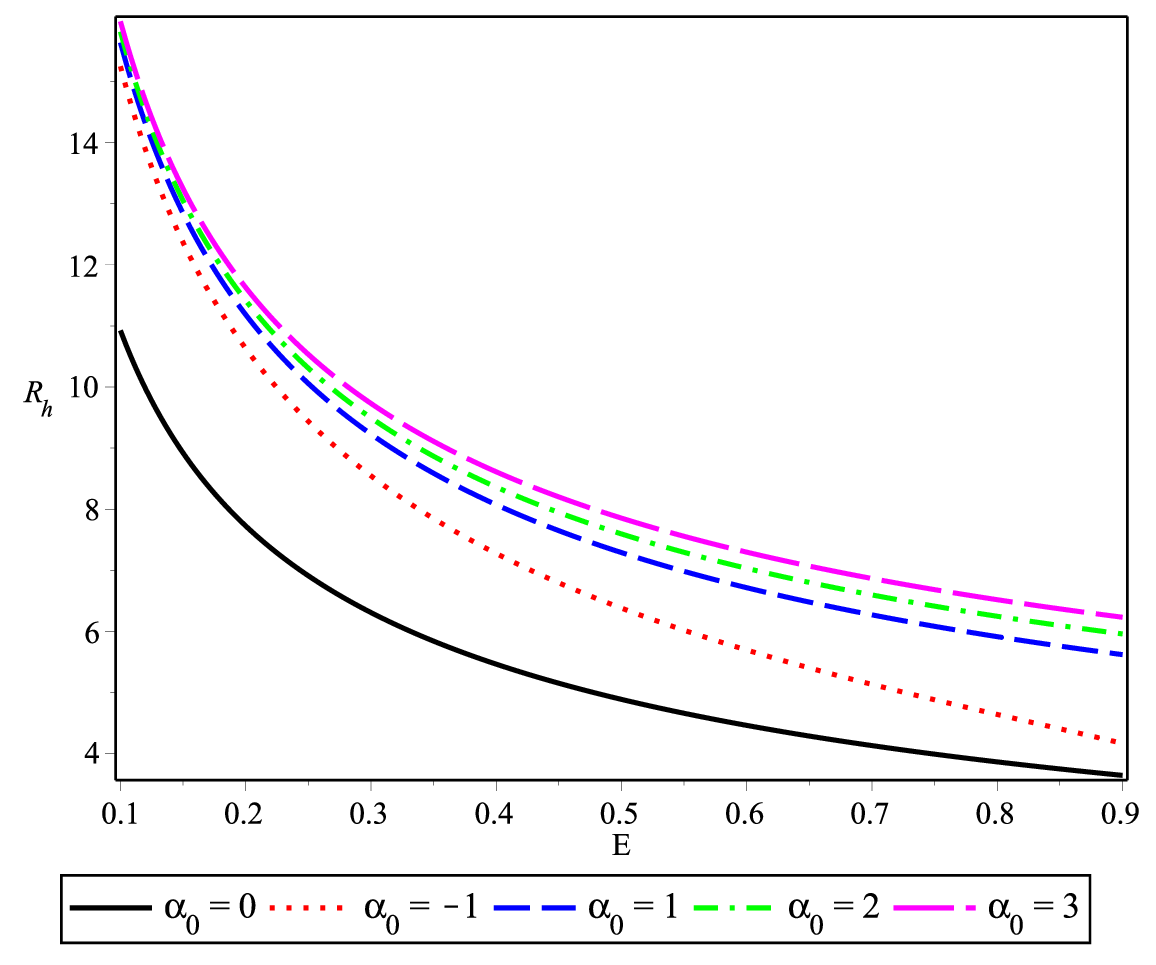}
	\caption{Horizon radius versus energy for $n=4$, $V=100$ and different values of the coupling constant $\alpha_0$. The black solid curve represents the case of Einstein's theory. 
    }
	\label{fig1}
\end{figure}

\begin{figure}
\includegraphics[scale=0.5]{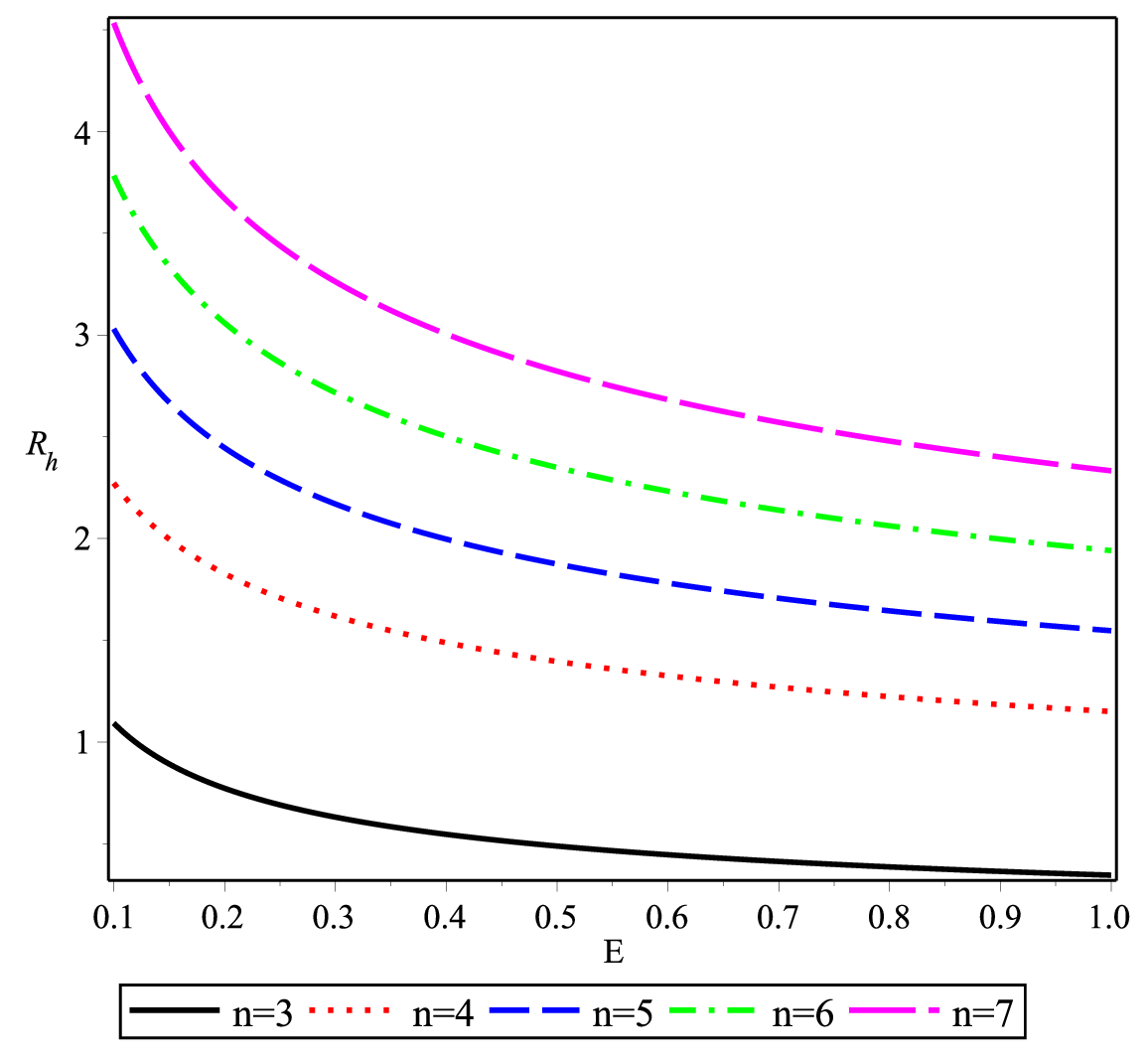}
	\caption{Horizon radius versus energy  for $\alpha_0 = 1$, $V=1$, and different values of the dimension $n$. The black solid line represents the case of Einstein's gravity.
       }
	\label{fig2}
\end{figure}

In Figs. \ref{fig1} and \ref{fig2}, we show the behavior of the horizon radius in terms of the energy for different values of the constants $\alpha_0$ and $n$. 
We see that, in fact, the radius of the FLRW apparent horizon in the EGB theory is always larger than in the case of Einstein's theory, and increases as the coupling constant and the number of dimensions increase.


%%%%%%%%%%%%%%%%%%%%%%%%%%
%%%%%%%%%%%%%%%%%%%%%%%%%%%

 \section{The laws of thermodynamics  for the FLRW apparent horizon}
\label{sec:laws}

According to classical thermodynamics, any thermodynamic system can be described by a fundamental equation from which all the properties of the system can be derived \cite{callen1998thermodynamics}. In this case,  the fundamental equation is given as the entropy of the apparent horizon (\ref{GBEntropy}) with the radius (\ref{GBrad}) and $\rho= \frac{E}{V}$. Then, we obtain explicitly 
\begin{equation}
S=\frac{n \Omega_n}{4} \left[\sqrt {{\frac {V}{
2\beta E} \left( 1+\sqrt {1+{\frac {4\alpha\beta E}{V}}} \right) }}
 \right] ^{n-1} \left[ 1+\frac {  4 ( n-1) \alpha\beta E}
 { ( n-3) V 
\left( 1+
\sqrt {1+{\frac{4\alpha\beta E
}{V}}} \right)}
 \right],
 \label{GBfeq}
\end{equation}
an expression that relates the entropy $S$ with the energy $E$ and the volume $V$ of the horizon.  
In the limit $\alpha_0=0$ and $n=3$,  this expression reduces to 
\begin{equation}
    S=\frac{3V}{8 E}, 
\end{equation}
which is the entropy of the FLRW apparent horizon in Einstein gravity \cite{sanchez2023thermodynamics}.

\begin{figure}
    \centering
\includegraphics[scale=0.5
]{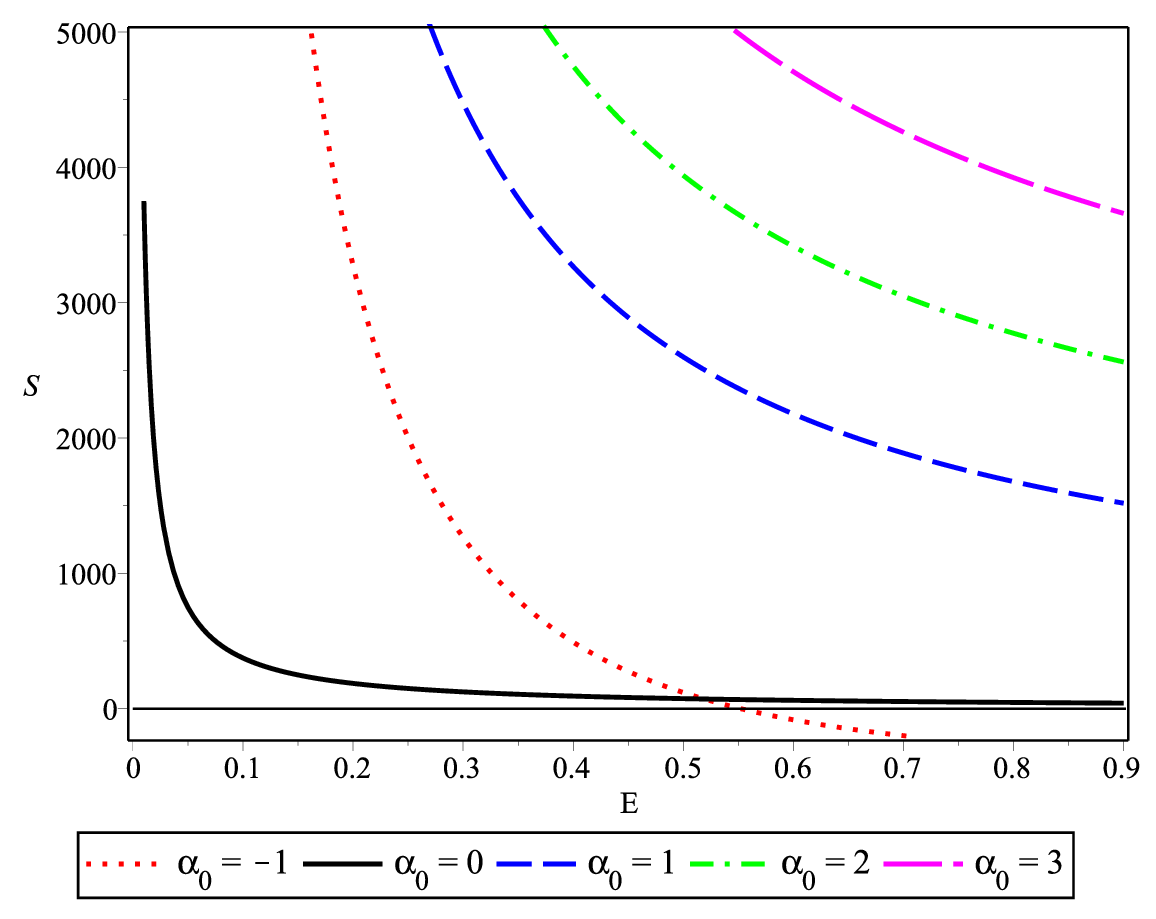}
    \caption{Entropy of the apparent horizon as a function of the energy $E$ for $V=100$, $n=4$, and different values of the coupling constant $\alpha_0$. The black solid line represents the entropy in Einstein's gravity.}
    \label{fig3}
\end{figure}

\begin{figure}
    \centering
\includegraphics[scale=0.5]{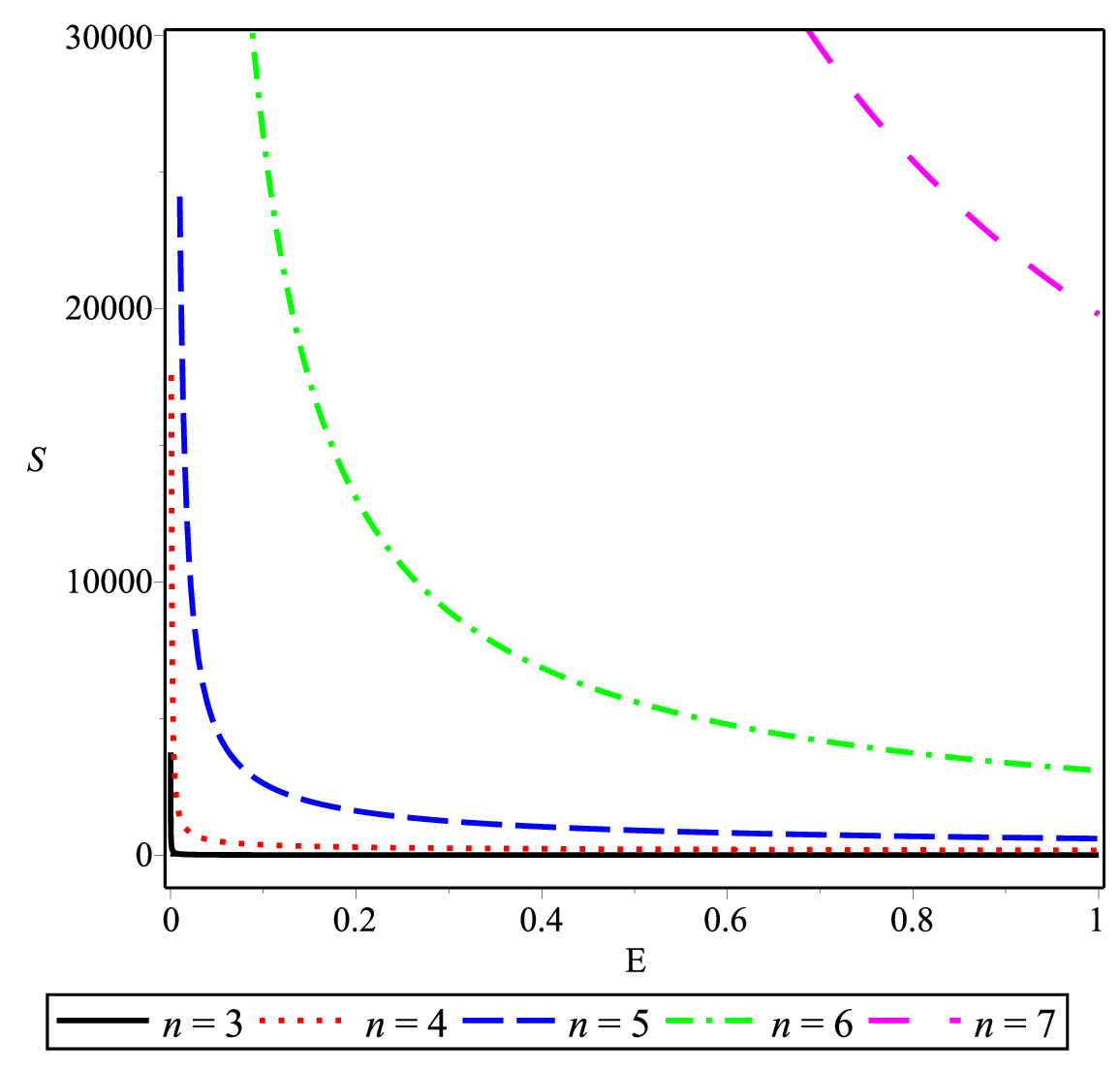}
    \caption{Entropy of the apparent horizon as a function of the energy for $V=1$, $\alpha_0=1$, and different values of the dimension parameter $n$.}
    \label{fig4}
\end{figure}

In Fig. \ref{fig3}, we illustrate the behavior of the entropy (\ref{GBfeq}) in terms of the energy for different values of the coupling constant. We see that for negative values of $\alpha_0$, the entropy can become negative, which we interpret as a non-physical behavior. In general, for positive values of $\alpha_0$ the entropy is always larger than in the case of Einstein gravity. Figure \ref{fig4} shows the entropy in terms of the energy for a fixed value of $\alpha_0$ and different dimensions. In all the cases, the entropy is larger than in Einstein's gravity.

%%%%%%%%%%%%%%%%%%%%

\subsection{The first law of thermodynamics}
\label{sec:flaw}

\begin{figure}
    \centering
\includegraphics[scale=0.5]{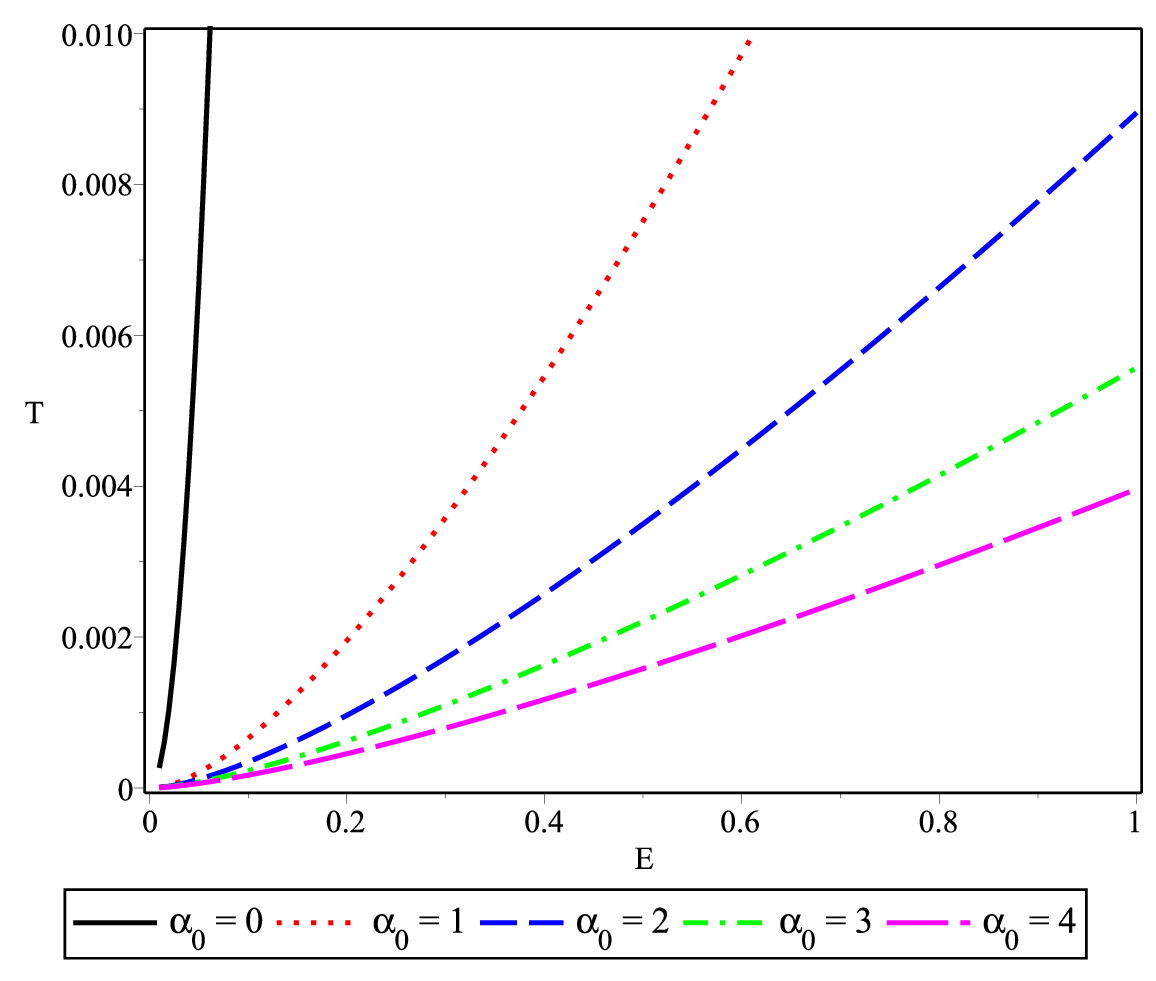}
    \caption{Temperatue of the apparent horizon as a function of the energy for $n=4$, $V=1$, and different values of the coupling constant $\alpha_0=1$. The black solid line represents the temperature in Einstein's gravity.}
    \label{fig5}
\end{figure}

In the case of an apparent horizon, the first law can be derived using the law of conservation of energy, which states that the change of internal energy is equivalent to the amount of energy that crosses the horizon during its evolution \cite{cai2005first}. Then, the first law can be expressed as
\be
dE = - T dS + W dV.
\ee
Then, the temperature and work density are determined as
\be 
T = - \left( \frac{\partial S}{\partial E} \right) ^{-1}
\label{temp1}
\ee
and
\be
W = - \frac{\partial S}{\partial V} 
\left( \frac{\partial S}{\partial E} \right) ^{-1},
\label{work1}
\ee
respectively. To compute the temperature, we use the relationship $\beta E/V = (R_h^2+\alpha)/R_h^4$ that follows from the expression for the radius of the apparent horizon (\ref{GBrad}). Then, the temperature can be expressed as 
\be
T = \frac{V}{2 \pi \Omega_n R_h^{n+1}},
\label{temp2}
\ee
which in the case $n=3$ leads to the temperature in Einstein's theory \cite{sanchez2023thermodynamics}
\begin{equation}
    T_E=\frac{8E^2}{3V}.
\end{equation}
In Fig. \ref{fig5}, we plot the temperature in terms of the energy for different values of the coupling parameter. In general, we observe that in the EGB theory the temperature of the apparent horizon is less than in Einstein's gravity.

Furthermore, according to Eq. (\ref{work1}), we obtain for the work density 
\begin{equation}
    W=\frac{E}{V} .
    \label{work2}
\end{equation}
Remarkably, the expression for the work density (\ref{work2}) is identical in both the EGB and the Einstein theories. We will see in Sec. \ref{sec:eos} that this leads to important consequences in the thermodynamic behavior of the apparent horizon in the EGB theory.

The above results for the thermodynamic variables can be used to compute the response functions \cite{callen1998thermodynamics}, which in this case are 
the heat capacity 
\be
C_V = T\left(\frac{\partial S}{\partial T}\right)_V =-{\frac {\Omega_n\,n \left( R_h^{2}+2\,\alpha \right)  \left( n-1 \right) R_h
^{n-3}}{4(n+1)}},
\ee
the compressibility 
\be
\kappa_T = - \frac{1}{V}\left(\frac{\partial V}{\partial p}\right)_T=-{\frac { 8 \pi\left( n+1 \right) R_h^{4} }{ \left( R_h^{2}+2\,\alpha
 \right) n \left( n-1 \right) }},
\ee
and the thermal coefficient
\be
\alpha_p = \frac{1}{V}\left(\frac{\partial V}{\partial T}\right)_p = \frac{1}{T} =
\frac{2 \pi \Omega_n R_h^{n+1}}{V}
.
\ee
All the response functions depend explicitly on the dimension number $n$ and the GB coupling constant $\alpha_0$. 
We see that in the Einstein limit, all these expressions correctly reduce to the results previously reported in \cite{sanchez2023thermodynamics}.
According to Eq.(\ref{condR}), $R_h^2+2\alpha >0$ and, therefore, the heat capacity and  compressibility are negative quantities, a
property that can be associated with the long-range character of the gravitational interaction of the apparent horizon  \cite{lynden1968gravo,lynden1980consequences}. 

\begin{figure}
    \centering
\includegraphics[scale= 0.5]{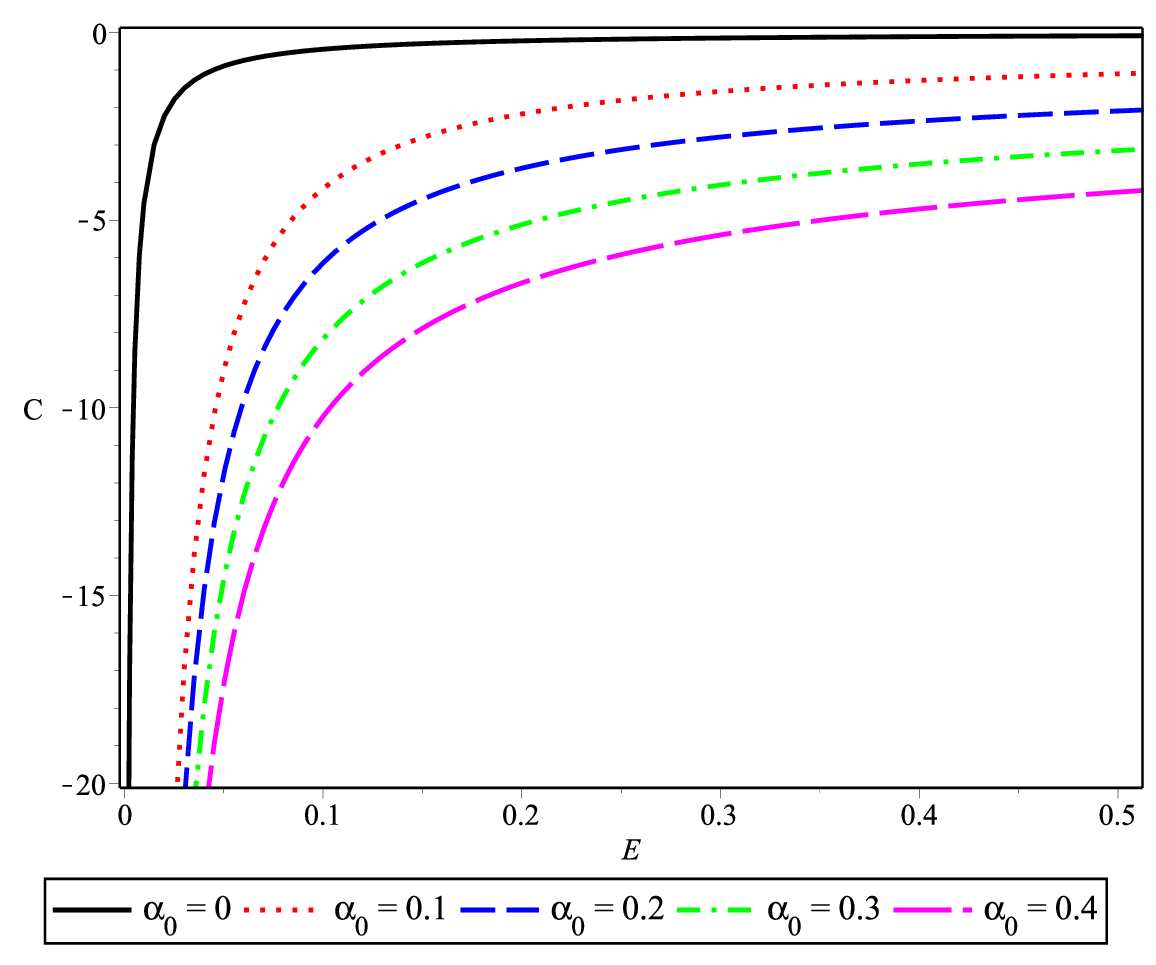}
    \caption{The heat capacity as a function of the energy of the apparent horizon. The solid black curve represents the Einstein limit. Here, $n=4$ and $V=1$ for concreteness.}
    \label{fig6}
\end{figure}

In Fig. \ref{fig6},
we illustrate the behavior of the heat capacity in terms of the energy for $n=4$ and different values of $\alpha$. 
We see that the heat capacity is, in fact, a negative quantity which decreases in value as the coupling constant increases.  

Moreover, using the relationship $C_p=C_V+\frac{TV\alpha_p^2}{\kappa_T}$, we get
\be
C_p=-{\Omega_n \frac { 4\,n\alpha\,  R_h^{n-3} \left( n-1 \right) 
 +  R_h^{n-1
} \left( 2\,{n}^{2}+n+3  \right)   }{8(n+1)}},
\ee
Consequently,  the adiabatic index can be expressed as
\be
\gamma= \frac{C_p}{C_V} = {\frac {  R_h^{2} \left( 2\,{n}^{2}+n+3 \right)  +4\,
\alpha\,n \left( n-1 \right) }{2 n \left( n-1 \right)  \left( R_h^{2}+2
\,\alpha \right) }},
\label{id_Cap}
\ee
an expression that reduces to $\gamma = 2$ in 
Einstein's theory.

On the other hand, from the relationship  $\kappa_s=\kappa_T-\frac{TV\alpha_p^2}{C_p}$, we obtain 
\be
\kappa_s =-{\frac { 24 \pi \left( n+1 \right) ^{2}R_h^{6}}{n \left( n-1 \right) \left( R_h^{2}+2\,
\alpha \right)   \left[R_h^{2} \left( 2\,{n}^{2}+
n+3 \right) +4\,\alpha\,n \left( n-1 \right)  \right] }},
\ee
and
\be
\frac{\kappa_T}{\kappa_s}= {\frac { R_h^{2} \left( 2\,{n}^2+n+3  \right) +4\,
\alpha\,n \left( n-1 \right) }{ 3 \left( n+1 \right) R_h^{2}}}.
\label{id_K}
\ee
It is then easy to see that
\be
\frac{C_p}{C_V} = \frac{3(n+1)R_h^2}{2n(n-1)(R_h^2+2\alpha)}\frac{\kappa_T}{\kappa_s}.
\ee
In classical thermodynamics, we have $\gamma=\frac{\kappa_T}{\kappa_s}=\frac{C_p}{C_V}$, which is satisfied only for $n=3$ and $\alpha=0$. We see that the thermodynamic relationships become generalized in EGB due to the presence of the coupling constant and the number of dimensions. The standard classical relationships turn out to be valid only in Einstein gravity.

\subsection{The second law}
\label{sec:slaw}

To analyze the second law in a consistent way, it is necessary to take into account that the apparent horizon is surrounded by matter that should also contribute to the entropy balance \cite{bekenstein1974generalized,bokhari2010generalized,saridakis2021generalized}.
 So, the total entropy $S_T$ consists of the geometric entropy $S$ and the matter entropy $S_m$. As for the geometric entropy, its derivative with respect to time can be computed from Eq.(\ref{GBEntropy}) yielding  
\begin{equation}
    \dot{S} = \frac{n\Omega_n  \left( n-1 \right)}{4} R_h^{n-4} \left(R_h^{2}+2\,\alpha \right)  \dot{R}_h.
    \label{sdot}
\end{equation}
Replacing here Eq.(\ref{dotR}), we obtain
\begin{equation}
    \dot{S}= \pi n \rho \Omega_n   H R_h^{n+1}
 \left( \omega+1 \right).
 \label{dotS}
\end{equation}
Notice that the geometric entropy alone does not satisfy the second law in general, but only for $\omega\geq -1$, which would imply a limitation of the applicability of classical thermodynamics to the FLRW apparent horizon. This shows the importance of considering the generalized second law, which takes into account the entropy of the entire system.

To determine the entropy of the matter, we consider the first law in the form
\begin{equation}
    TdS_m = dE_m + p dV,
\end{equation}
where $E_m=V \rho$ with $V=\Omega_n R_h^n$. Then, using the conservation law in the form $d\rho =-n(\omega+1) \rho H dt$, we obtain
\begin{equation}
    TdS_m = \rho (\omega+1) n \Omega_n R_h^{(n-1)}dR_h - \Omega_n R_h^n n( \omega+1)\rho H dt ,
\end{equation}  
an expression that using $T=\frac{1}{2 \pi R_h}$  reduces to 
\begin{equation}
  \dot{S}_m =  2 \pi n \Omega_n \rho (\omega+1)  (\dot{R}_h- H R_h) R_h^n.
  \label{dotSm}
\end{equation}

Finally, adding the two contributions, that is, the change in the entropy of the apparent horizon $\dot{S}$ and the change in the entropy of the matter $\dot{S}_m$, we obtain the change in the total entropy $\dot{S}_T = \dot{S} + \dot{S}_m$ as
\begin{equation}
   \dot{S}_T = {\frac {16 
\Omega_n  n {\pi }^{2}\left( \omega+1
 \right) ^{2}\rho^{2}H R_h ^{n+5}  }{ \left( n-1 \right)  \left( R_h
 ^{2}+2\,\alpha \right) .}}
\end{equation}
Taking into account  condition (\ref{condR}), the above result shows that the change of the total entropy is always positive in accordance with the generalized second law of thermodynamics.

%%%%%%%%%%%%%%%%%%%%%%%%

\section{Equation of state of the apparent horizon}
\label{sec:eos}

According to Eq.(\ref{work0}), in general, 
the work density is related to the pressure and the energy density by $W=(\rho-p)/2$. On the other hand, in the case under consideration,  the fulfillment of the first law implies Eq.(\ref{work2}), i.e., $W=\rho$. Consequently, from the fulfillment of these two independent conditions for the work density, it follows that the apparent horizon as a thermodynamic system must satisfy the equation of state 
\be
p+\rho=0,
\label{eos}
\ee
i.e., it can be associated with a 
dark energy fluid. Consequently, according to Eqs.(\ref{dotS}) and (\ref{dotSm}), the entropy of the apparent horizon and matter are constant quantities separately, indicating that the evolution of the apparent horizon corresponds thermodynamically to an isentropic process. Moreover, the dark-energy equation of state (\ref{eos})
 is valid independently of the number of dimensions and the GB coupling constant, indicating a sort of thermodynamic universality of the apparent horizon in gravity metric theories based only on curvature. 

\begin{figure}
    \centering
\includegraphics[scale=0.6]{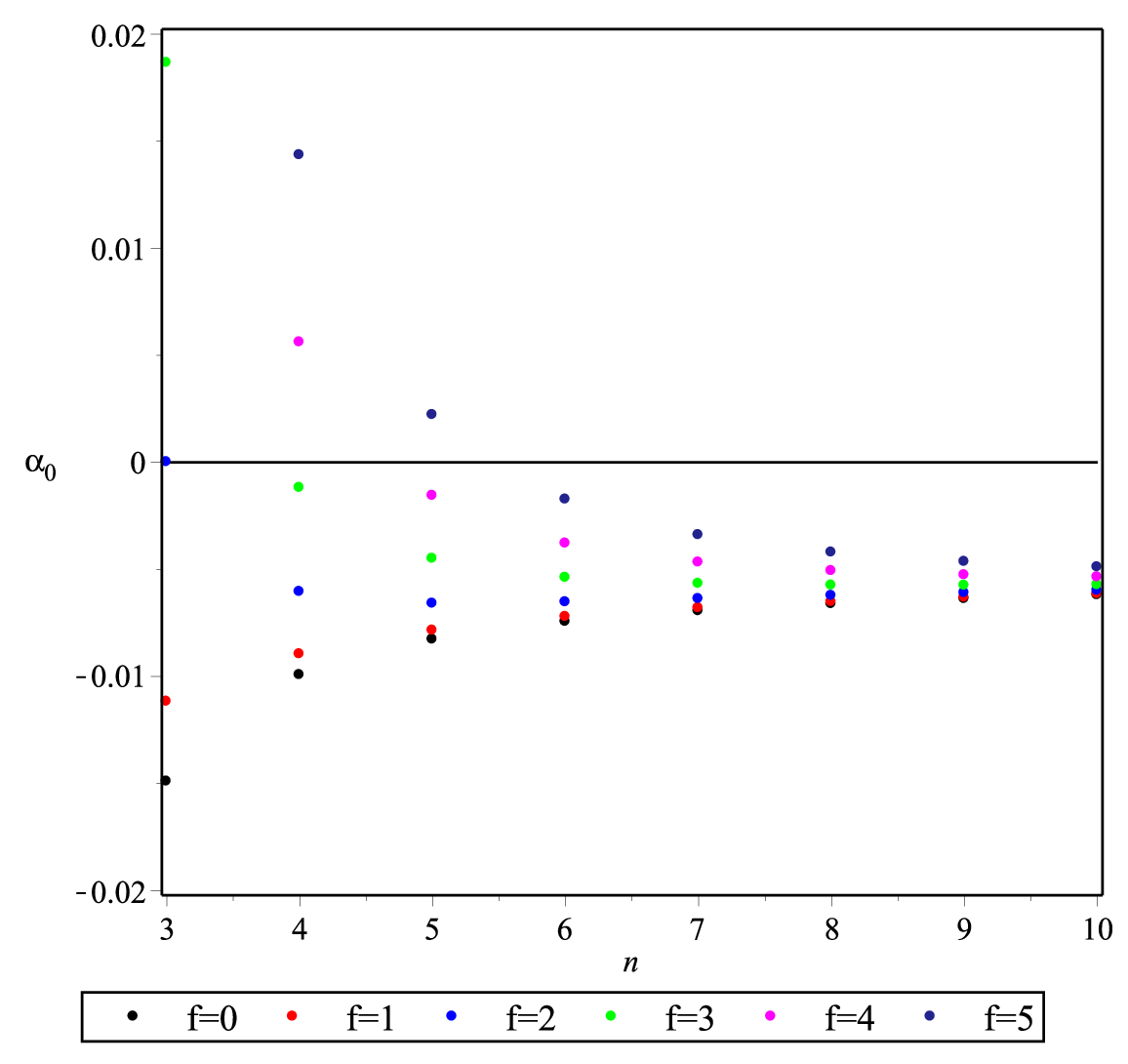}
    \caption{The GB coupling constant as a function of the number of dimensions $n$ for different values of the number of degrees of freedom $f$. Here we set $\rho_0 =1.$}
    \label{fig7}
\end{figure}

Furthermore, the response functions are also constant as well as the adiabatic index (\ref{id_Cap}), a behavior that is usually associated with ideal gases. On the other hand, from the equipartition theorem it follows that the adiabatic index of ideal gases should be related to the number of thermodynamically  accessible degrees of freedom $f$ of the system by \cite{callen1998thermodynamics}
\be
\gamma = 1 + \frac{2}{f}.
\label{adindex}
\ee
In the case of Einstein gravity, we obtain $f=2$, which is less than the number of degrees of freedom of an ideal gas, a result that can be interpreted as due to the long-range character of gravity \cite{sanchez2023thermodynamics}. Then, since $f$ represents the number of degrees of freedom, it seems reasonable to require $f$ to be an integer. Using Eqs.(\ref{id_Cap}) and (\ref{adindex}), it is then straightforward to show that the following relationship holds
\be
\alpha_0 = \frac{n}{64\pi (n-2)\rho_0}\left[
\frac{9(n+1)^2}{16n^2(n-1)^2}f^2 -1 
\right].
\label{condgamma}
\ee
This is an interesting equation that relates explicitly the GB coupling constant with the number of degrees of freedom of the apparent horizon and implies a discretization of the EGB theory at the level of the action. In fact, for any given integer value of $f$, the coupling constant can assume only a finite number of values determined by the value of $n$. We illustrate this result in Fig. \ref{fig7}.
The values of $\alpha_0$ are discrete and depend on the number of dimensions $n$ and degrees of freedom $f$. Even the limiting value $f=0$ is allowed and interestingly, according to Eq.(\ref{cond1}), corresponds to the minimum value of the entropy. For a fixed value of $f$, the coupling constant decreases as $n$ increases and tends asymptotically to the value of minimum entropy with $\alpha_0 \to -\frac{1}{64\pi \rho_0}$. For the values $f=0,1,2$, where $f=2$ corresponds to Einstein's theory, only negative values of the coupling constant are allowed. However, for higher values, $f\geq 3$, the coupling constant can become positive for small values of $n$.

\section{Conclusions}
\label{sec:con}

In this work, we explored the thermodynamic properties of the apparent horizon of the FLRW spacetime in EGB theory. It was shown that the main thermodynamic variables depend explicitly on the dimension number and the GB coupling constant. From this point of view, it can be argued that the properties of the FLRW apparent horizon in the EGB theory are different from those of Einstein gravity.

However, by imposing the laws of thermodynamics on the evolution of the apparent horizon, it turns out that along the entire evolution a dark-energy-like equation of state is satisfied, independently of the values of the coupling constant and the dimension number. This proves a universal behavior of the apparent horizon in all metric theories in which the action depends only on the curvature. Moreover, it turns out that the thermodynamic evolution is isentropic and the heat capacities are constant as in the case of ideal gases. 

In addition, we argue that from the relation between the adiabatic index and the number of thermodynamically accessible degrees of freedom, one can obtain an expression that relates the GB coupling constant with the dimension number and the number of degrees of freedom, which we assume as an integer. As a result, the GB coupling constant becomes a discrete quantity, pointing out a discretization of the EGB theory at the level of the action.

%%%%%%%%%%%%%%%%%%%%%%%%%%%%%
%%%%%%%%%%%%%%%%%%%%%%%%%%%%

\section*{Acknowledgments}

This work was partially supported by UNAM-DGAPA-PAPIIT, grant No. 108225, and Conahcyt, grant No. CBF-2025-I-243.

%\bibliography{references}
%\end{document}


\end{document}